\documentstyle[12pt]{article}
\title{Free Energies of
Isolated 5- and 7-fold Disclinations in Hexatic
Membranes}

\author{Michael W.\ Deem and David R.\ Nelson\\
Lyman Laboratory of Physics\\Harvard University, Cambridge, MA\ \ 02138}

\begin{document}
\maketitle
\renewcommand{\baselinestretch}{1.3}
\tiny
\normalsize

\begin{center}
{\bf{Abstract}}
\end{center}

We examine the shapes and energies of 5- and 7-fold disclinations
in low-temperature hexatic membranes.  These defects buckle at different
values of the ratio of the bending rigidity, $\kappa$, to the hexatic
stiffness constant, $K_A$, suggesting {\em two} distinct
Kosterlitz-Thouless defect proliferation temperatures.
Seven-fold disclinations are studied in detail numerically for
arbitrary $\kappa/K_A$.
  We argue that thermal
fluctuations always drive $\kappa/K_A$ into an ``unbuckled'' regime
at long wavelengths, so that disclinations should, in fact,
proliferate at the {\em same} critical temperature.
 We show analytically that both types
of defects have power law shapes with continuously variable
exponents in the ``unbuckled'' regime.  Thermal fluctuations
then lock in specific power laws at long wavelengths, which
we calculate for 5- and 7-fold defects at low temperatures.

\begin{center} PACS 05.70.Jk, 68.10.-m, 87.22.Bt \end{center}
\newpage

\section{Introduction}

     Topological defects, such as dislocations and disclinations,
behave differently in membranes and monolayers.
Monolayers are films, typically with a triangular
lattice in their low-temperature crystalline phase,
 strongly confined to a plane by, for example, surface
tension.  In monolayers, point-like topological imperfections are the
ingredients of a detailed two-stage melting theory,\cite{1,2} which predicts
that the usual latent heat associated with a first order melting
transition can be spread out over an intermediate hexatic phase
characterized by long range bond orientational order and short
range positional order.\cite{2,3}   In contrast to monolayers,
membranes can easily buckle out of the two-dimensional plane.  This
buckling
lowers defect energies.  The standard example of a membrane is an
extended lipid bilayer surface in water solution.\cite{4} Dislocation
energies in membranes are finite, in contrast to a
logarithmic divergence with system size for  monolayers, which leads to
melting of crystalline membranes at any nonzero temperature for
entropic reasons.\cite{NelsonIII,SeungII} The generic low temperature phase for
membranes at large enough length scales is inevitably a hexatic
liquid with long range bond orientational order and a vanishing
shear modulus.

     Disclinations can also lower their energy by buckling.  The
energy of a single disclination in an otherwise crystalline
monolayer diverges with the square of the system size. Buckling in
a membrane leads to energies of plus and minus disclinations  that
diverge only logarithmically with size.\cite{NelsonIII} Interestingly, the
coefficients of these logarithmic divergences are different: the
energy of 5-fold (plus) disclinations is about a factor of 2 lower
than that of a 7-fold (minus) ones.\cite{SeungII}
  This asymmetry differs from the
behavior of defects in most monolayer materials that exhibit the
Kosterlitz-Thouless defect unbinding transition.\cite{2} Plus and minus
vortices in superfluid helium films, for example, must have exactly
the same core energies and logarithmic divergences with system size
by time reversal invariance.  Equality of energies also holds for
dislocations with equal and opposite Burgers vectors in crystalline
monolayers.  Plus and minus disclinations in hexatic monolayers
have different  core energies, due to the different local
environments of the 5- and 7-fold defects, but the coefficients of
their logarithmically diverging energies are identical.\cite{1}
Identical coefficients for the logarithmically diverging $\pm$ defect
pairs ensures that the elementary Kosterlitz-Thouless energy-entropy
 balance leads to the same proliferation temperature for
each type of defect.  This famous argument \cite{Kosterlitz}
predicts that 5- or 7-fold
defects proliferate whenever the free energy to create a disclination, $F_5(R)
= E_5(R) - 2 k_{\rm B} T \ln(R/a_0)$ or
$F_7(R) = E_7(R) - 2 k_{\rm B} T \ln(R/a_0)$,
becomes negative.  Here
$E_5(R)$ and $E_7(R)$ are disclination energies as a function of
the system size $R$, and $a_0$ is a microscopic cutoff.

     The disparate disclination energies in buckled crystalline
membranes suggests that these energies may also differ in hexatic
membranes, as recently emphasized
 by one of us.\cite{NelsonII} The energy of 5-fold
disclinations in hexatic membranes has been studied by
Seung,\cite{Seung} Guitter,\cite{Guitter} and Guitter and
 Kardar.\cite{GuitterII} The  results
depend on two dimensionless parameters,
 $\kappa/k_{\rm B} T$ and $K_A/k_{\rm B} T$,
where
$\kappa$ and $K_A$ are the membrane bending rigidity and hexatic stiffness
constant, respectively.  When  $\kappa/K_A \gg 1$, the membrane remains
asymptotically flat in the presence of both 5- and 7-fold defects,
and the  energies diverge logarithmically with equal coefficients.
When  $\kappa/K_A < 11/72$, however, the 5-fold disclination buckles and
reduces its in-plane bond angle energy at the expense of bending
energy.  The resulting energy still diverges logarithmically, but
the coefficient is reduced by buckling.  The locus of
Kosterlitz-Thouless transition temperatures for 5-fold defects in membranes
when buckling is included is shown by the solid curve in Fig. 1.
These disclinations assume a conical shape when  $\kappa/K_A < 11/72$ and, as
shown in Sec.\ 3,
exhibit a weaker power law deviation from flatness
when $\kappa/K_A > 11/72$, with continuously
variable exponents.

     In this paper, we study 7-fold defects and
 determine the curve for their
proliferation, shown as the dashed curve in Fig. 1.  Because
buckled 7-fold defects do not have the azimuthal symmetry of 5-fold
defects, some numerical work is necessary.  As anticipated in
\cite{NelsonII}, the 7-fold hexatic disclination energy exceeds its 5-fold
counterpart for  $\kappa/K_A \ll 1$, leading to a larger Kosterlitz-Thouless
critical temperature.  We find that 7-fold disclinations buckle to reduce
the coefficient of their logarithmically diverging energy whenever
$\kappa /K_A < 13/216$, while they remain asymptotically
flat when  $\kappa/K_A$ exceeds this  value.
There is again a power law deviation from flatness for $\kappa/K_A > 13/216$.
As is evident from Fig. 1, naive application of the Kosterlitz-Thouless
criterion to hexatic membranes leads to {\em two} distinct
defect proliferation temperatures over a significant range of
parameters.

     Our results provide useful information about deformations of
hexatic membranes near defects at relatively low temperatures.
However, understanding the behavior at very large distances from
the defect cores requires that we take into account the nonlinear
renormalization of  $\kappa$ and $K_A$ by thermal fluctuations.  These effects
were first studied in a perturbative, low-temperature expansion by
F.\ David {\em et al.},\cite{DavidII}
 who found that the hexatic phase of membranes
is controlled by a line of fixed points in the
 $(  k_{\rm B} T / K_A , k_{\rm B} T / \kappa )$ plane
with slope  $\kappa^*/K_A^* = 1/4$.
The parameters $\kappa$ and $K_A$ are driven toward the fixed line by
thermal fluctuations even if they initially lie in one of the buckled regimes.
  This line and the associated nearby
renormalization group flows are indicated by the heavy curve in Fig. 1.
The unstable Kosterlitz-Thouless fixed line
for hexatic monolayers ($\kappa \to \infty$)
is also shown.
Because the stable fixed line has a slope far below the critical slopes
for buckling of 5- or 7-fold
disclinations, we conclude that these defects probably
have symmetrically diverging energies when thermal fluctuations are
taken into account.  Provided that the line of stable fixed points,
which is only known perturbatively at low temperatures,\cite{DavidII}
 does not bend
significantly before piercing the vertical part of the
Kosterlitz-Thouless instability curve, both
defects will remain unbuckled and proliferate at the same point.
Although the coefficients of the logarithmic divergences in 5-
and 7-fold disclination energies are the same, these defects have
interesting power law shapes in the ``unbuckled'' regime.

     Park and Lubensky have recently incorporated fluctuations in the
internal metric of the membrane into the work of David {\em et al.}\cite{Park}
Inclusion of these fluctuations appears to be necessary to account for
local in-plane shear and
compression modes, whose primary effect is to dilate and
reshuffle the
nearest-neighbor bond connectivity of the atomic or molecular
constituents.  Averaging over these modes leads to an {\em effective}
coupling $K_A^{\rm eff}$  which should replace the
hexatic stiffness in the model used here.
The bending rigidity is unchanged.
  The remaining
renormalization of $\kappa $ and $K_A^{\rm eff}$
 by thermal undulations, however, is identical
to that found by David {\em et al.}\cite{DavidII}
 Thus, our overall conclusions are
unchanged, provided we use $\kappa$ and the renormalized
coupling $K_A^{\rm eff}$
 in our results and in Fig.\ 1. In
agreement with the results of the study of disclination
statistical mechanics by Park and Lubensky,\cite{Park} we still
expect a
symmetrical $\pm$ disclination unbinding transition with unbuckled
defects in hexatic
membranes.  Disclinations cannot unbind separately, in contrast to
the predictions of the ``naive'' Kosterlitz-Thouless argument,
provided the thermal renormalization of the ratio
  $\kappa/K_A^{\rm eff}$ to $1/4$ at long wavelengths remains intact
out to the transition temperature.\cite{Note1}

In independent work, Park and Lubensky have also studied the buckling
of 7-fold disclinations.\cite{ParkII}  Their result for the critical
ratio $\kappa/K_A$ and the behavior near the buckling transition
are in agreement with the results presented here.  Our results are
more reliable in the limit $\kappa/K_A \gg 1$, however.

     It is worth noting that similar issues involving disclination
asymmetry arise even for monolayers, when two-dimensional nematic
order is present.  The topologically stable defects are now  $\pm 1/2$
disclinations, and the ordered state is described by both bend and
splay elastic constants $K_1$ and $K_3$.\cite{13}
 When $K_1 \ne   K_3$, the energies of
$\pm$ disclinations again diverge logarithmically with different
coefficients.  Here, a renormalization group analysis of nonlinear
effects due to thermal fluctuations shows that $K_1$ and $K_3$ are driven
to equality at long wavelengths, so that the ``one-Frank-constant
approximation'' becomes asymptotically correct at long
wavelengths.\cite{14} The disclinations energies are equal in this
limit, and one again expects a symmetrical defect unbinding
transition.\cite{15}

     We should stress that even though disclination energies are
asymptotically symmetrical in hexatic membranes,
buckling will still occur locally for appropriate parameter
values.  Buckling will persist out to length scales such that
the renormalized value of the ratio $\kappa/K_A^{\rm eff}$ is
in the unbuckled regime.
An intrinsic
microscopic asymmetry, moreover, can still arise in the
liquid, after the disclinations unbind.  As emphasized in Ref.\ \cite{Seung},
a net excess of disclinations should have important consequences in
liquid membranes with free boundary conditions  or a
topology that can change on experimental time scales.  Exactly how
the $\pm$ disclination populations become identical as one approaches
the liquid-to-hexatic transition from the liquid phase is an
interesting topic for future research.

In Sec.\ 2 we discuss how
membrane buckling can screen disclinations.  In Sec.\ 3 we
review analytical calculations of the energy of a 5-fold
disclination.  We use approximate theory and exact numerical
calculations to calculate the energy and shape of a 7-fold disclination in
Sec.\ 4.  We review the important
 effects of thermal fluctuations, including how
these fix the asymptotic defect shapes in the unbuckled regime, in
Sec.\ 5.

\section{Curved Hexatic Membranes with Defects}

The Hamiltonian for a flexible, hexatic membrane is given in the
limit of vanishing surface tension by
\cite{NelsonIII,David}
\begin{eqnarray}
H &=& H_\theta + H_\kappa + H_{\bar \kappa} \nonumber \\
H_\theta &=& \frac{K_A}{2} \int d^2 S (\partial_i \theta -
\Omega_i) g^{ij} (\partial_j \theta - \Omega_j) \nonumber \\
H_\kappa &=& \frac {\kappa}{2} \int d^2 S {\bar H}^2 \nonumber \\
H_{\bar \kappa} &=& \bar \kappa \int d^2 S K \ .
\label{1}
\end{eqnarray}
All these integrals are over the surface of the membrane.  For the
case of a membrane with free
 boundary conditions there should also be line tension and
geodesic curvature terms.   We neglect these terms.  Here
$g^{ij}$ is the inverse metric tensor, $\bar H$ is the mean
curvature, $K$ is the Gaussian curvature, and the gauge field is
defined by $K = \gamma^{ij} D_i \Omega_j$.  Upon defining $g = \det
(g_{ij})$, we have $\gamma^{ij} = \epsilon_{ij} g ^{-1/2}$ with
$\epsilon_{11} = \epsilon_{22} = 0$ and $\epsilon_{12} =
-\epsilon_{21} = 1$.  The surface area element is given by $d^2 S =
d^2 \sigma g^{1/2}$.

We are interested in very flexible membranes, as opposed to monolayers,
 and so we will neglect the possible
surface tension term of the form $H_r = r \int d^2 S$.
  We will also ignore the Gaussian
curvature term $H_{\bar \kappa}$, which is a perfect derivative by the
Gauss-Bonnet theorem.
$H_\kappa$ is the standard
bending energy term, and  $H_\theta$ is the contribution to the energy
from fluctuations in the local bond order parameter.  The bond order
parameter is frustrated by the rotation of tangent vectors that
occurs under parallel transport on a curved surfaces.  The amount of
frustration is given by the gauge field, $\Omega_i$.

To gain some physical understanding of a flexible hexatic membrane,
we examine the ground states.  In particular, we search for the
low temperature
geometries of 5- and 7-fold disclinations.  After performing the
minimization over $\theta$, we find
\begin{equation}
\left. \frac{\delta H }{\delta \theta(\sigma)}
 \right\vert_{\theta = \theta_0}
= g^{-1/2} \partial_j \left[g^{ij} (\partial_i \theta_0 - \Omega_i)\right]
 = 0 \ ,
\label{2}
\end{equation}
where $\theta_0$ is the bond angle field that minimizes the energy.
Upon defining
\begin{equation}
\partial_i \theta_0 - \Omega_i = \gamma_i^{~j} \partial_j \chi \ ,
\label{3}
\end{equation}
we find the derivative ({2}) is zero except when derivatives of
$\chi$ do not commute.  We can show that disclinations cause the
derivatives to fail to commute by applying the operator $\gamma^{ki}
\partial_k$ to Eq.\ \ref{3}:\cite{NelsonII}
\begin{eqnarray}
D_i D^i \chi &=& K - \gamma^{ki} \partial_k \partial_i \theta_0 \nonumber \\
&=& K(\sigma) - s(\sigma) \nonumber \\
&\equiv& c(\sigma) \ .
\label{4}
\end{eqnarray}
The disclination density is given by
\begin{equation}
s(\sigma) = \sum_i s_i \delta(\sigma - \sigma_i) g(\sigma_i)^{-1/2} \ .
\label{5}
\end{equation}
Here the disclination strength is given by $s_i = \pi/3$ for a 5-fold
disclination and $s_i =-\pi/3$ for a 7-fold disclination.  Given
the form of Eq.\ (\ref{4}), we can express $\chi$ in terms of a
Green's function as
\begin{equation}
\chi(\sigma) = \int d^2 S' G(\sigma \vert \sigma') c(\sigma') \ ,
\label{6}
\end{equation}
where
\begin{equation}
D_i D^i G(\sigma \vert \sigma') = \nabla_\sigma^2 G(\sigma \vert \sigma') =
\delta(\sigma - \sigma') g(\sigma')^{-1/2} \ .
\label{7}
\end{equation}
The Hamiltonian when the bond angle field is minimized is given by
\begin{equation}
H_\theta = \frac{K_A}{2} \int d^2 S d^2 S' d^2 S''
c(\sigma') c(\sigma'')
\left[\partial_i G(\sigma \vert \sigma')\right] g^{ij}
\left[\partial_j G(\sigma \vert \sigma'')\right] \ .
\label{8}
\end{equation}

We see Eq.\ (\ref{8}) that the relevant
quantity is not the disclination density or the Gaussian curvature
separately, but
 rather the difference, $c(\sigma)$, between them.  Consequently,
the hexatic energy arising from a disclination can be reduced by a
non-zero Gaussian curvature.  This screening, of course, will cost
the membrane in terms of bending energy.  The competition between
screening of the hexatic energy and bending energy determines the
equilibrium shape of the membrane.

   The energy of a single, isolated disclination with ``charge''
 $s$ in a flat, circular
membrane is given by $E = ( s^2 / 4 \pi) \ln (R/a_0)$.\cite{1}
Here $R$ is the radius of the membrane and $a_0$ is a microscopic
cutoff.  For 5- and 7-fold disclinations, $E_5 = E_7 = (\pi K_A / 36)
\ln (R/a_0)$.  Buckling of the membrane can reduce this energy.
We describe the location of the membrane by
\begin{equation}
{\bf X}(r,\phi) = (r \sin \phi, r \cos \phi, f(r, \phi)) \ .
\label{9}
\end{equation}
The diverging contribution to the  energy
comes from the large $r$ region of the surface.
The bending energy can diverge no more strongly than
$\log (R/a_0)$ in
a buckled ground state, since  otherwise the energy would
{\em increase} upon buckling.
This bound implies that $f$ grows at most linearly with $r$.
If $f$ grows {\em less} rapidly than $r$, then the Green's function defined
by
Eq.\ (\ref{7}) is given by $\partial_r G \sim 1/(2 \pi r)$ as
$r \to \infty$.\cite{Note2}  Furthermore,
the Gauss-Bonnet theorem then implies that $\int d^2 S K = 0$ for membranes
with a disk-like topology.\cite{Coxeter}
  From Eq.\ (\ref{8}), we see that the hexatic energy remains
$E \sim (\pi K_A / 36) \ln (R/a_0)$  and has not been reduced.
  For the logarithmic hexatic energy to be screened
by buckling, therefore, the height must grow linearly with r:
\begin{equation}
{\bf X}(r,\phi) = (r \sin \phi, r \cos \phi, r h(\phi)) \ .
\label{9a}
\end{equation}

The Green's function that satisfies Eq.\ (\ref{7}) is then given by
\begin{equation}
G(r,\phi) = b^{-1} \ln \left[r(1 + h(\phi)^2)^{1/2} \right] \ ,
\label{10}
\end{equation}
with
\begin{equation}
b = \int d \phi \frac{(1 + h^2 + h'^2)^{1/2}}{1 + h^2} \ .
\label{11}
\end{equation}

To evaluate the hexatic Hamiltonian, we need both the Gaussian
and mean curvature.
For the surface defined by Eq.\ (\ref{9}), the Gaussian curvature
is proportional to a delta function:
\begin{equation}
K(\sigma) = \alpha \delta(\sigma) g^{-1/2} \ .
\label{12}
\end{equation}
The coefficient, $\alpha$, can be determined from
\begin{eqnarray}
\alpha &=& \int d^2 S K  \nonumber \\
&=& \int d^2 S \gamma^{ij} \partial_i \Omega_j \nonumber \\
&=& \int_c d \sigma^i \Omega_i \ ,
\label{13}
\end{eqnarray}
where $c$ is a contour bounding the surface.
If ${\bf e}_1$ and ${\bf e}_2$ are an orthonormal basis for vectors
tangent to the surface, the gauge field is given by \cite{David}
\begin{equation}
\Omega_i = {\bf e}_1 \cdot \partial_i {\bf e}_2 \ .
\label{14}
\end{equation}
{}From this equation we find, by  taking ${\bf e}_1$ and ${\bf e}_2$ to be
basis vectors in polar coordinates and subracting the result
for a flat surface,
\begin{eqnarray}
\Omega_1 &=& 0 \nonumber \\
\Omega_2 &=& 1 - \frac{(1 + h^2 + h'^2)^{1/2}}{1 + h^2} \ .
\label{15}
\end{eqnarray}
We, therefore, conclude that
\begin{equation}
\alpha = 2 \pi - b \ .
\label{16}
\end{equation}
We can now perform the integrals in Eq.\ (\ref{8}) to find
\begin{equation}
H_\theta = \frac{K_A (2 \pi - s - b)^2}{2 b} \ln (R/a_0) \ .
\label{17}
\end{equation}
The mean curvature is given by \cite{Coxeter}
\begin{eqnarray}
\bar H &=& \nabla \cdot \frac {\nabla f}{(1 + \vert \nabla f \vert^2)^{1/2}}
 \nonumber \\
&=& \frac{(h + h'')(1 + h^2)}{r (1 + h^2 + h'^2)^{3/2}} \ .
\label{18}
\end{eqnarray}
The bending energy is, then, given by
\begin{equation}
H_\kappa = \frac{\kappa}{2} \ln (R/a_0) \int d \phi
\frac{(h + h'')^2(1 + h^2)^2}{(1 + h^2 + h'^2)^{5/2}} \ .
\label{19}
\end{equation}
The contribution to the bending energy associated with
the singularity at $r=0$ will be absorbed into a core energy.
The total energy of a hexatic membrane with a single, isolated disclination,
excluding the core contribution, is given by
\begin{equation}
\frac{H}{\ln (R/a_0)} =
K_A \frac{(2 \pi - s - b)^2}{2 b} +
\frac{\kappa}{2} \int d \phi
\frac{(h + h'')^2(1 + h^2)^2}{(1 + h^2 + h'^2)^{5/2}} \ .
\label{20}
\end{equation}
The geometry of lowest energy is found by minimizing with respect
to the function $h(\phi)$.  Note that $b$ depends on $h(\phi)$
though Eq.\ (\ref{11}).

The route from the covariant Hamiltonian (\ref{1}) to the tractable
expression (\ref{20}) is complicated.  For a nearly flat surface,
a simplified Hamiltonian is often used \cite{NelsonIII,NelsonII}
\begin{equation}
H = \frac{K_A}{2} \int d^2 r \left[ \partial_i \theta - A_i \right]^2 +
 \frac{\kappa}{2} \int d^2 r (\nabla^2 f)^2 \ ,
\label{21}
\end{equation}
where
\begin{equation}
A_i = \frac{1}{2}
\epsilon_{jk} \partial_k \left[ (\partial_i f) (\partial_j f)\right] \ ,
\label{22}
\end{equation}
and the derivatives are in flat space.
  The bond angle field, $\theta_0(\sigma)$ that
minimizes this energy is given by
\begin{equation}
\partial_i (\partial_i \theta_0 - A_i) = 0 \ .
\label{23}
\end{equation}
To satisfy this equation, we define
\begin{equation}
(\partial_i \theta_0 - A_i) = \epsilon_{ij} \partial_j \chi \ .
\label{24}
\end{equation}
Applying the operator $\epsilon_{ik} \partial_k$ to this
equation, we find \cite{Seung}
\begin{equation}
\nabla^2 \chi = (\partial_x^2 f)(\partial_y^2 f)
-(\partial_x \partial_y f)^2 - s({\bf r}) \ ,
\label{25}
\end{equation}
with
\begin{equation}
s({\bf r}) = \sum_i s_i \delta({\bf r} - {\bf r}_i) \ .
\label{25a}
\end{equation}
For an isolated 5- or 7-fold disclination,
$s({\bf r}) = \pm (\pi / 3) \delta({\bf r})$.
The Hamiltonian now reduces to
\begin{equation}
H = \frac{K_A}{2} \int d^2 r \vert \nabla \chi \vert^2
+ \frac{\kappa}{2} \int d^2 r (\nabla^2 f)^2 \ .
\label{26}
\end{equation}
We can further find the height function, $f$, which minimizes this
Hamiltonian.  It satisfies a second nonlinear, hexatic ``von Karmon
equation'':\cite{Seung}
\begin{equation}
\frac{\kappa}{K_A} \nabla^4 f =
(\partial_y^2 \chi)(\partial_x^2 f) +
(\partial_x^2 \chi)(\partial_y^2 f) -
2(\partial_x \partial_y \chi) (\partial_x \partial_y f) \ .
\label{27}
\end{equation}
With the simple Hamiltonian (\ref{21}), then, we have explicit
partial differential equations that define the surface of minimal
energy.  For the covariant Hamiltonian, the differential equation
arising from minimizing Eq.\ (\ref{20}) is much more complex.

\section{The Energy of a 5-fold Disclination}
A 5-fold disclination can be screened by  a surface with a positive
Gaussian curvature.  The natural surface to consider is a cone.

We first review the results of the approximate Hamiltonian
(\ref{21}).  A cone defined by $f(r) = a r$ solves Eqs.\ (\ref{25})
and (\ref{27}), with $\chi(r) = - (\kappa/K_A) \ln (r/a_0)$.\cite{Seung}
 The coefficient is given by $a^2 = 1/3 - 2 \kappa /
K_A$.  For $\kappa/K_A < 1/6$, the membrane buckles.  The energy is
given by
\begin{equation}
E_5 \approx \left\{
\begin{array}{l}
(\pi \kappa / 3)(1 - 3 \kappa / K_A) \ln (R/a_0),~
     \kappa / K_A < 1/6 \\[.2in]
(\pi K_A / 36) \ln (R/a_0) ,~\kappa / K_A > 1/6
\end{array}  \right. \ .
\label{28}
\end{equation}

We now review the results of the covariant Hamiltonian (\ref{1}).\cite{Guitter}
  Equation (\ref{20}) fully specifies the energy, with
$h=a$ and $b = 2 \pi (1 + a^2)^{-1/2}$.  We first note that Eq.\
(\ref{16}) can be derived from a geometrical argument.  We consider
capping off the cone with a small sphere of radius $\epsilon$, as in
Fig.\ 2.   The bending energy is unaffected by this small
perturbation, since we are ignoring the contribution near $r=0$.
The Gaussian curvature is zero everywhere except on the
sphere.  On the sphere it is given by $\alpha = \int d^2 S K = 2 \pi
\int_{\cos \psi}^1 d u = 2 \pi (1- \cos \psi)$.    With $\tan \psi =
a$, we have $\alpha = 2 \pi [1 - (1 + a^2)^{-1/2}]$, in agreement with
Eq.\ (\ref{16}).  Upon defining $x = (1 + a^2)^{-1/2}$, we have
\begin{equation}
\frac{H}{\pi K_A \ln (R/a_0)} = \frac{(1/6 - 1 + x)^2}{x} +
 \frac{\kappa}{K_A} \frac{1-x^2}{x} \ .
\label{29}
\end{equation}
Minimization of this equation leads to
\begin{equation}
E_5 = \left\{
\begin{array}{l}
\pi K_A \left\{ 2 \left[ (25/36 + \kappa/K_A)(1-\kappa/K_A) \right]^{1/2}
 - 5/3 \right\}\\
{}~~~~~~~~~~~~~~~ \times \ln (R/a_0),~
     \kappa / K_A < 11/72 \\[.2in]
(\pi K_A / 36) \ln (R/a_0) ,~\kappa / K_A > 11/72
\end{array}  \right. \ .
\label{30}
\end{equation}
We note that the limit $K_A \to \infty$ corresponds to the
inextensional limit of a crystalline membrane.  This energy has the
correct limit $E_5 \to (11 \pi \kappa / 30) \ln (R/a_0)$ as $K_A \to
\infty$, which corresponds to a {\em crystalline} membrane.\cite{SeungII}
  While the Hamiltonian (\ref{21}) is often
thought of as valid for small $\nabla f$, we see that it does not
exactly predict the buckling transition, where $\nabla f$ is a small, nonzero
constant.

When $\kappa / K_A > 11/72$, the above calculation shows that the height
grows sublinearly with $r$.  In fact, we now show that the height
grows with a power that depends continuously on $\kappa / K_A$.  We assume
that $f(r,\phi) = f(r)$.  The hexatic and bending energies of Eq.\ (\ref{1})
are then given by
\begin{eqnarray}
F_\theta &=& \pi K_A \int_1^\infty  dr \frac{(1 + f'^2)^{1/2}}{r}
\left[\frac{1}{(1 + f'^2)^{1/2}} - \frac{5}{6}\right]^2 \nonumber \\
F_\kappa &=& \pi \kappa \int_1^\infty dr \frac{ (1 + f'^2)^{1/2}}{r}
\left[\frac{f'}{(1 + f'^2)^{1/2}} +
\frac{r f''}{(1 + f'^2)^{3/2}} \right]^2 \ .
\label{30a}
\end{eqnarray}
We have set the short-range cutoff to $a_0 = 1$.
As we shall see, $f' \to 0$ as $r \to \infty$, when
$\kappa / K_A > 11/72$.
Upon expanding Eq.\ (\ref{30a}) for small $f'$, we find
\begin{equation}
F \sim \pi \int_1^\infty dr \left[
\frac{K_A}{36 r} + \frac{\kappa(f' + r f'')^2}{r}
- \frac{11 K_A f'^2}{72 r} \right] + O(f^4) \ .
\label{30b}
\end{equation}
Upon solving the equation $\delta F / \delta f(r) = 0$,
 we find in the case of constant moduli
\begin{equation}
f(r) \sim a r^{1 - y} {\rm ~~as~~} r \to \infty  \ ,
\label{30c}
\end{equation}
with
\begin{equation}
y = [1 - 11 K_A / (72 \kappa)]^{1/2} \ .
\label{30d}
\end{equation}
We also minimize Eq.\ (\ref{30a}) numerically.
  We express
$f'(r)$  on a grid at grid points $r_i = \exp(\ln r_{\rm max} i/n)$ and
approximate
the integral by a sum and derivatives by finite difference.
We found convergence was achieved for $n=200, r_{\rm max}=100$.  Results
are presented for $n=400, r_{\rm max}=200$.
The Polak-Ribiere conjugate gradient method was used
to determine $f'(r_i)$.\cite{Press}  Figure 3 shows the
height as a function of $r$ for the specific case
$\kappa/K_A = 1/4$.  The numerical results reproduce the asymptotic
scaling of Eq.\ (\ref{30c}).  The energy of this ground state is
$E_5 = (\pi K_A / 36) \ln R -
 0.0281 \pi K_A + E_{\rm c}$, where $E_{\rm c}$ is a
core energy contribution.  If this core contribution is sufficiently large,
$E_{\rm c} > 0.0281 \pi K_A$, the surface of minimal energy would be flat, and
the constant $a$ in Eq.\ (\ref{30c}) would be zero.

\section{The Energy of a 7-fold Disclination}
A 7-fold disclination can be screened by  a surface with a negative
Gaussian curvature.  There is no obvious natural surface to consider
in this case.
  Using the approximate Hamiltonian (\ref{21}), we can achieve an
analytical answer, however.  We let
\begin{eqnarray}
f(r,\phi) &=& a r \sin 2 \phi \nonumber \\
\chi(r) &=& \frac{3 \kappa}{K_A} \ln (r/a_0) \ .
\label{31}
\end{eqnarray}
Equation (\ref{27}) is solved by this choice.  Equation (\ref{25}) is
solved provided $a^2 = 2/9 - 4 \kappa / K_A$.  When $\kappa/K_A < 1/18$,
the membrane buckles.  The energy is given by
\begin{equation}
E_7 \approx \left\{
\begin{array}{l}
(\pi \kappa )(1 - 9 \kappa / K_A) \ln (R/a_0),~
     \kappa / K_A < 1/18 \\[.2in]
(\pi K_A / 36) \ln (R/a_0) ,~\kappa / K_A > 1/18
\end{array}  \right. \ .
\label{32}
\end{equation}

The covariant Hamiltonian (\ref{1}) does not yield so easily
to an analytical treatment.  We can, however, expand Eq.\ (\ref{20})
for small $h$ to find the buckling transition:
\begin{equation}
\frac{\delta H }{\delta h(\phi)} = 0 = \frac {13 K_A}{72}(h + h'')
+ \kappa (h + 2 h'' + h'''') + O(h^3) \ .
\label{32a}
\end{equation}
This equation predicts buckling for $\kappa/K_A < 13/216$ with
$h(\phi) = a \sin 2 \phi$.
Again the approximate Hamiltonian (\ref{21})
predicts the transition value only approximately.

We determine the surface that
minimizes the energy for general values of $\kappa/K_A$
by numerically identifying the function
$h(\phi)$ that minimizes Eq.\ (\ref{20}).
  We express
$h(\phi)$  on a grid at grid points $\phi_i = 2 \pi i / n$, again approximating
the integral by a sum and derivatives by finite difference.
We found convergence was achieved for $n=100$.  Results are presented
for $n=200$.  The Polak-Ribiere conjugate gradient method was used
to determine the $h(\phi_i)$.\cite{Press}

Figure 4 presents the numerically-determined energies $H_\theta$,
$H_\kappa$, and $H$.  The membrane buckles when
$\kappa / K_A < 0.060$, in good agreement with the exact value
of $13/216$
As the membrane becomes less stiff, the
buckling is able to screen more and more of the hexatic energy:
the hexatic energy goes from
\begin{equation}
H_\theta \sim
 (\pi K_A / 36) \ln (R/a_0) ~~~~~{\rm for}~~~~~ \kappa/K_A \ge 13/216,
\label{35}
\end{equation}
to
\begin{equation}
H_\theta \sim 0 ~~~~~{\rm as}~~~~~ \kappa/K_A \to 0,
\label{36}
\end{equation}
as expected.
Similarly, the bending
energy goes from
\begin{equation}
H_\kappa \sim 0 ~~~~~{\rm for}~~~~~ \kappa/K_A  \ge 13/216,
\label{37}
\end{equation}
to
\begin{equation}
H_\kappa \sim 2.27 \kappa \ln (R/a_0)
 ~~~~~{\rm as}~~~~~ \kappa/K_A \to 0 \ .
\label{38}
\end{equation}
The limit as $ \kappa/K_A \to 0$ agrees with
numerical calculations for inextensional crystalline membranes.\cite{SeungII}

Figure 5 shows the function $h(\phi)$ for various values of the
ratio $\kappa/K_A$.  As expected, the surface is flatter for stiffer surfaces.
For very flexible membranes, the surface converges to a limiting shape.
This limiting shape is very nearly proportional to $\sin 2 \phi$,
as shown in Figure 6.
More generally, we can expand $f(r,\phi)$ in a Fourier series
\begin{equation}
f(r,\phi) \ \sum_{m=0}^{\infty} f_m(r) \cos(m \phi) \ ,
\label{38aa}
\end{equation}
where $f_m(0) = 0$.
All odd terms must vanish for a two-fold symmetric saddle point
configuration.  In addition, $f$ should change sign under a $\pi/2$ rotation,
which eliminates the terms in Eq.\ (\ref{38aa}) with $m=0, 4, 8, \ldots$.
Such a symmetric saddle has an expansion of the form
\begin{equation}
f(r,\phi) \ \sum_{p=0}^{\infty} f_{4p+2}(r) \cos\left[2(2 p+1) \phi\right] \ ,
\label{38aaa}
\end{equation}
a conclusion also reached by Park and Lubensky.\cite{ParkII}.
We have checked numerically that the only non-zero Fourier components in
Eq.\ (\ref{38aa}) are indeed of the form $m=4p+2$, although
the $m=2$ term alone provides an excellent approximation.

When $\kappa / K_A > 13/216$, the height
grows sublinearly with $r$.   Just as for the 5-fold disclination,
the height
grows with a power that depends continuously on $\kappa / K_A$.
To see this, note first than
when $\kappa / K_A > 13/216$, $\nabla f \to 0$ as
$r \to \infty$.  Upon expanding Eq.\ (\ref{1}) for small $\nabla f$, we find
\begin{eqnarray}
F &\sim& \frac{K_A}{144} \int dr d\phi ~\left[
2/ r + 13 (\partial_r f)^2/r  + 11 (\partial_\phi f)^2/r^3 - 12
\left(\partial r (\partial_\phi f)^2\right)/r^2 \right]
\nonumber \\
&+& \frac{\kappa}{2} \int dr d\phi ~r\left[\partial_r^2 f + (\partial_r f)/r
+ (\partial_\phi^2 f)/r^2\right]^2 + O(f^4) \ .
\label{38a}
\end{eqnarray}
The solution of $\delta F / \delta f(r,\phi) = 0$ is
\begin{equation}
f(r) \sim a r^{1-y} \sin 2 \phi {\rm ~~as~~} r \to \infty  \ ,
\label{38b}
\end{equation}
with
\begin{equation}
y = \left[720 + 13 K_A/\kappa -
(331776 + 29952 K_A/\kappa + 169 K_A^2/\kappa^2)^{1/2}\right]^{1/2} / 12 \ .
\label{38bb}
\end{equation}
As in the case of 5-fold disclinations, this sublinear decay
leads to an additive, constant correction to the logarithmically
diverging energy as $R \to \infty$.

\section{Thermal Fluctuations}
We have so far ignored thermal fluctuations of the hexatic membrane.
This assumption
 is valid only for the $T \to 0$ limit.  For finite temperatures,
and for large membranes, thermal fluctuations will become important.

A complete discussion of thermal effects is beyond the scope of this
paper.  We can, however, use the results of David {\em et al.} \cite{DavidII}
and of Park and Lubensky \cite{Park} to estimate how the structure of
disclinations in hexatic membranes is modified at finite temperatures.
Park and Lubensky argue that proper implementation of an ultraviolet
cutoff to fluctuations in hexatic membranes leads to the replacement
\begin{equation}
K_A \to K_A^{\rm eff} =
 K_A - \frac{3}{32 \pi} k_{\rm B} T (K_A / \kappa)^2 \ .
\label{39a}
\end{equation}
The bending rigidity $\kappa$ is unchanged.
 In the absence
of a non-zero disclination density, the remaining renormalization
equations for $\kappa$ and $ K_A^{\rm eff}$ are those found originally
by David {\em et al.}:\cite{DavidII}
\begin{eqnarray}
\frac{d  K_A^{\rm eff}} {d l} &=& 0 \nonumber \\
\frac{d \kappa / k_{\rm B} T} {d l} &=& - \frac{3 }{4 \pi}\left (
1 - \frac{ K_A^{\rm eff}}{4 \kappa}
\right) \ .
\label{39b}
\end{eqnarray}
The renormalization group flows induced by these equations are indicated
schematically by the arrows in Fig.\ 1.

We apply these results to the dilute limit of isolated disclinations discussed
in Secs.\ 4 and 5. The locus of disclination unbinding transitions,
given by the criteria $F_5(\kappa, K_A) \equiv 0$ and
$F_7(\kappa, K_A) \equiv 0$ discussed in the Introduction, are shown as the
solid and dasked lines in Fig.\ 1.
When thermal fluctuations are superimposed on
the solutions of the $T=0$ extremal equations for disclinations in a membrane
of size $R$, standard finite size scaling arguments suggest that the couplings
controlling the defect energies on this scale should be the {\em running}
coupling constants $ K_A^{\rm eff}(l)$ and $\kappa(l)$ obtained from Eq.\
(\ref{39b}) evaluated at $l=\ln(R/a_0)$.
Effects of thermally-excited bound disclination pairs on an otherwise
isolated defect could be included by adding
a vortex fugacity to the set of recursion
relations.\cite{Park}
We then expect that Eqs.\ (\ref{30a}) and (\ref{38a}) should be
replaced by expressions where $K_A$ and $\kappa$ are replaced by the functions
$K_A^{\rm eff}(l = \ln r/a_0)$ and $\kappa(l = \ln r/a_0)$ appearing
inside the integrals over $r$.  Although $K_A^{\rm eff}$ does not renormalize
at this order, a nontrivial renormalization could appear when higher order
corrections in $k_{\rm B} T/\kappa$ and $k_{\rm B} T/K_A^{\rm eff}$
are included.

If the basin of attraction of the locally stable fixed line in Fig.\ 1
includes the entire hexatic phase, $K_A^{\rm eff}(r)$ and $\kappa(r)$
will always be driven as $r \to \infty$ into the unbuckled regime for
both 5- and 7-fold disclinations, since $ \lim_{r \to \infty} \kappa(r) /
K_A^{\rm eff}(r) = 1/4$.  The bending energy will then not contribute
to the logarithmically diverging part of the energy, and both
defects should unbind at the same point.  Although ``unbuckled'' in this
sense, the limited disclination shapes will be characterized by the power laws
(\ref{30c}) and (\ref{38b}), with $\kappa/K_A = 1/4$.
For 5-fold defects we find the asymptotic shape is given by
$y=y_+ = 0.6236$, while for
7-fold defects we have $y=y_- = 0.8249$.

\bibliographystyle{new}

\section*{Figure Captions}

\flushleft{Figure 1.}
The phase diagram for proliferation of isolated 5-fold (solid) and
7-fold (dashed) disclinations.  Five- and seven-fold disclinations
buckle above lines (not shown) extending from the origin to the tops of
the vertical portions of the solid and dashed curves, respectively.
For small $\kappa$ or $K_A$
(outside the curves), disclinations proliferate.  Lines of
renormalized effective rigidities are also indicated
(bold).
Renormalization group flows obtained by David {\em et al.}\ in a low
temperature perturbation expansion away from the
unstable fixed line at $\kappa = \infty$ to the stable line describing
the crinked phase are indicated by the arrows.

\flushleft{Figure 2.} The cone $f(r) = a r$ shown in projection
 capped by  a
small sphere used to calculate the integrated Gaussian curvature.
  The angle $\psi$ is given by $\tan \psi = a$.

\flushleft{Figure 3.}
The height of an ``unbuckled'' membrane with a 5-fold disclination
as a function of $r$ for the case
$\kappa/K_A = 1/4$.  There is a short distance cutoff so that $f(r)$ is
undefined for $r < a_0$.

\flushleft{Figure 4.}
The hexatic (short dashed), bending (long dashed) and total (solid)
energies when $r \to \infty$ as a function of $\kappa / K_A$ for
a 7-fold disclination.  An overall factor of $\ln (R/a_0)$ has
been suppressed in each term.  The defect is
unbuckled for $\kappa / K_A > 13/216 \approx 0.060$.

\flushleft{Figure 5.}
The surfaces $h(\phi)$
for a 7-fold defect above its unbuckling transition
for the cases $\kappa / K_A$ = 0.06, 0.05, 0.03,
0.01, and 0.001.

\flushleft{Figure 6.}
The surface $h(\phi)$ in the limit $\kappa / K_A \to 0$ (solid), which mimics
the behavior in a crystalline solid,  and
the function $0.534 \sin(2 \phi)$ (dashed) for a 7-fold defect.

\end{document}